# The COVID19 'infodemic' – the role and place of academics in science communication

**Jennifer Cole, PhD**

**Department of Geography,**
Royal Holloway University of London,
Egham Hill,
Egham,
Surrey TW20 0EX
United Kingdom

Jennifer.cole@rhul.ac.uk

## Abstract

As the COVID19 pandemic has spread across the world, a concurrent pandemic of information has spread with it. Deemed an 'infodemic' by the World Health Organization, and described as an overabundance of information, some accurate, some not, that occurs during an epidemic, this proliferation of data, research and opinions provides both opportunities and challenges for academics. Academics and scientists have a key role to play in the solutions to the infodemic challenge: as educators, influences and communicators, even where their expertise and experience does not align precisely with the SARS-Cov2 virus and its impacts.

Successful communication requires a better understanding of how the public seeks, understands and processes scientific information, however, in order to maximise the ways in which experts engage with traditional and social media and to make sure that such engagement does not add to confusion and misinformation alongside efforts to counter or challenge it. This paper will outline the key advantages to be had from greater engagement with COVID19 discussions, the popular channels through which such discussions take place and through which information is disseminated. It also warns against the common pitfalls those who choose to engage might encounter, whilst stressing that the disadvantages of doing so are far outweighed by the advantages such engagement offers.

**Introduction**

As the COVID19 pandemic has spread across the world, a concurrent pandemic of information has spread with it [1]. Deemed an 'infodemic' by the World Health Organization, and described as, "an overabundance of information, some accurate, some not, that occurs during an epidemic" [2] this proliferation of data, research and opinions provides both opportunities and challenges for academics and scientists. Academics and scientists have key roles to play in the solutions to these challenges.

In March 2020, WHO Director-General Tedros Adhanom Ghebreyesus called for the need to manage the COVID19 infodemic [3]. The tendency is to approach the infodemic as challenging, and to focus predominantly on the misinformation and disinformation that spreads through informal rumour networks, social and traditional media [4] – and to see only the challenges. The flow of information also offers opportunities for embracing knowledge exchange and dissemination, however: but this depends on a better understanding of the ways in which information is exchanged. This, in turn, is dependent on how that information fits with communicative ecologies – the socio-cultural framing and analysis of the local context in which communication occurs and which shapes its meaning [5]. Such opportunities are particularly important to embrace as, in the digital age, information can spread even more quickly than viruses [6], and so public health messages that can help with disease prevention and control can be disseminated quickly, as can messages that can counter misinformation before it is able to take hold. Information – both good and bad – spreads through populations in similar ways to the spread of disease [7], potentially offering opportunities to ensure information on protective behaviours reaches at-risk populations ahead of the virus. WHO has identified five key roles in infodemic management: scan and verify evidence; explain the science; amplify the reach of messages; measure the infodemic and assess trends and impacts; and co-ordinate and govern information technology [1]. As a key stakeholder in the production, interpretation and dissemination of knowledge, the academic sector has a particularly important role to play in the first three of these.

As educators are well aware, people enjoy sharing information: the role of social diffusion in information spread is well documented [8, 9]. However, this is true even when an individual does not believe the information they are sharing [10]. Disinformation (information passed on maliciously when it is known to be false, leading to a defective information process [11]) and misinformation [12] (false information shared in the mistaken belief that it is valid) far predate social media [13], for example as far back in history as the 'Black Death' of the 14[th] century, false rumours included accusations that the disease was due to Jews poisoning wells and streams [14]. Fake news has abounded through a wide variety of media and across political history [15] – including many US elections [16]. Rumours and falsehoods spread through whatever media people use to communicate: verbal, written, broadcast or digital. So too does the reliable information that can counter, correct and complete those falsehoods – as long as those who hold better information can mobilise themselves quickly and efficiently enough.

Early pioneers of communication science talked of communication as bits of information whose flow could be quantified between sender(s) and receiver(s) [17], and academics since the earliest days of the internet have considered the importance of focussing on the social construction of knowledge as well as its content [18]. More than 30 years ago, Pierre Lévy wrote about the internet as a new kind of knowledge space [19] – as transformative to knowledge flows as the development of language, notation such as writing, mathematics and cartography, and the invention of printing – but he also warned that the key challenge for humanity in the computer age was to understand how to order and process this knowledge, so that we do not simply "haul masses of information around with us" (p10). We should be heedful of this during the current pandemic: by November 2020, there were more than 76,000 academic publications on COVID19 indexed on PubMed (not including pre-prints), with new ones appearing every day – all attracting interest from lay public and media as well as the academy.

Cyberspace is an enabler that allows for the discovery, publication and application of new knowledge, the dissemination of best practices and information, and the exchange of views and opinions. It thus creates economic opportunities and a more educated workforce [20]. On the negative side, the internet also enables the spread of fake news [21], conspiracy theories [22] and misinformation [23], though often not to as great an extent that is often assumed [24, 25]. In the remainder of this paper, I will focus on key ways in which the academic community can help to combat the COVID19 infodemic: firstly by understanding what information people are looking for; by realising that even an imperfect response may be the best available; and by considering how non-academics receive information; and then by engaging with traditional and social media to deliver it.

**1. Understand what information people want and where they seek it.**

During a public health emergency, people seek out more information, more actively, and from more sources, than they did beforehand [26] in order to better understand the disease and the risks it poses to them. In short, people want to learn more about the disease – something which academics, as educators, can certainly help with. Academics fulfil many of the requirements for source credibility that helps information to be believed and acted on [27].

Dashboards that track location, number and rate of increase of cases in close to real time have proved particularly popular during COVID19 and previous outbreaks. Examples include the COVID19 Dashboard produced by Johns Hopkins University [28] and Worldometer [29], as well as those presenting future projections, such as the Institute of Health Metrics and Evaluation (IHME) [30]. Academic institutions, particularly those that already monitor and collect population health data, are ideally placed to create such data trackers quickly and ensure they are populated with accurate information.

How people process such information, differs, however: some prefer to read and self-educate, while others prefer to ask questions and discuss answers in a more socially interactive environment [31, 32]. Some people struggle to process complex concepts; they may be insufficiently practiced at critical thinking [33], and vulnerable to taking sensational newspaper headlines at face value [34]. As a result, academics and experts should consider not only what information they provide, but how clearly this is explained – particularly to those who may not have a good grounding in basic science or mathematics – and how it can be disseminated across many platforms simultaneously, including those with which academics may be less familiar. This in turn requires an understanding of the communicative ecology of the population that needs to be informed.

To reach people who prefer to learn by discussion and interaction, responding to FAQs (frequently asked questions, or question and answer sessions) on popular platforms, including TV shows, radio shows, newspapers and websites can be highly beneficial. The desire to ask questions increases the closer the risk comes and more the immediate individual perceives the risk to be [26]. The granularity of the answers required also increases. Academics contributing to FAQ sessions can enable more context-specific questions to be answered – regarding the risks associated with certain types of workplace [35], for example, or the risks to those with certain underlying health conditions. Identifying platforms through which such engagement can be provided, be those university websites, popular radio shows, AMA (Ask Me Anything) sessions on platforms such as [www.reddit.com](www.reddit.com) or local newspapers will help academics to engage with local communities. Academics may also be able to anticipate what questions community members benefit from having answered but may not initially think to ask [36] and can provide further explanation of terms and data that not may be easily understood [27] or which may have been widely misunderstood.

Previous research has shown that people rarely consult only one source of information: most consult several sources, across many different platforms, and are most likely to accept information when multiple sources agree [27]. They are more likely to turn first to familiar platforms they already use and trust, and only look further if these do not provide the information they require. As most laypeople do not routinely check university websites or follow academics' Twitter feeds, this may require academics to actively seek out trusted disseminators and go to the platforms the public use, rather than expect the public to come to them. In doing this, is it worth considering that media – whether traditional or social – is not 'good' or 'bad', high quality or low: each medium is unique and shaped by the information it mediates, meaning that the better the quality of the information it is given, the better the quality of the information it can disseminate. Individuals evaluate multiple sources and do not see 'social media' as a single type of information source but differentiate between messages passed by friends and family and those from official sources and expert voices [37].

It is worth being mindful that while scientists are widely trusted to understand the facts [38], the public may have less confidence in their ability to make political decisions based on these facts [39], however. Therefore, it may help academics to consider what type of expert the public will consider them to be and where their opinion will carry the most weight.

**2. Be prepared to go outside of your comfort zone (but not too far).**

A second consideration when engaging with traditional or media or social media during the COVID19 pandemic – and perhaps in apparent contradiction to the points made above – is that while academic research generally moves slowly and can take years to publish, with research findings presented to the expert community only after rigorous checking and re-checking of evidence, during the COVID19 pandemic no-one may have the equivalent expertise of a scientist who has spent a lifetime studying influenza, or HIV, or malaria. The best guess of an expert virologist, even if they are not a SARS-Cov2 expert, may be the best option available: do not be afraid to offer what expertise you have. This can be uncomfortable for academics but conspiracy theorists [40], anti-vaccination supporters [41], lockdown sceptics [42] and mask opposers [43] will not show the same restraint and misinformation thrives in a vacuum. During a fast-moving pandemic the luxury of waiting until one is 100 per cent sure before publishing a proof or discussing a hypothesis does not exist: the WHO infodemic training leans heavily on Voltaire's famous aphorism, already popular within healthcare: "don't let perfect be the enemy of the good" [44]. You may not know everything, but you will know enough to help the public to understand which mutations may change the game, and which won't; why case fatality rate estimates are likely to be much higher at the start of an epidemic than once better data becomes available; and why vaccine trials often pause for additional safety checks. Be prepared to say what you do know, even if it's not everything, and to explain why uncertainty at stages of a scientific process does not indicate failure. When a single expert is not available, the next best option is to ask a crowd [45] that can formulate an answer between its constituent members, drawing together their 'distributed' intelligence [19] and leading to consensus and agreement across a number of sources [46, 47]. This method was widely used by the American military during the Cold War, when the RAND Corporation developed the Delphi Method [48], a means by which an expert group – all of whom knew part of the answer – could collate and consolidate that knowledge when considering how the world might react to a nuclear war.

During the 2014-15 West Africa Ebola outbreak, people expressed frustration when there was a delay in the release of information and updates from official sources [31]. They would turn to less trusted sources to fill the gap – but would back-check this against the more reliable sources later, when those became available [26]. A guess from an expert is more likely to be correct than one from elsewhere but is only available if the expert is prepared to make it. Management theory often presents the importance of speaking quickly, but of being brave enough to retract if later proven wrong [49]: academics are less used to making such fast-moving decisions but may have a lot to learn from other fields where such practices are commonplace. Another lesson, from development studies, is how to present, evaluate and decide between a number of sub-optimal options when the most preferred choice is not available [50]. If you are not sure, and the chances are that no-one else is either, be willing to share what you do know. Even if it is incomplete and uncertain, it may be the best available.

**3. Consider how research will be interpreted by a non-academic audience**

During a pandemic, interest in science and scientific publications increases dramatically. People understandably want to know where the virus came from, how it spreads – due to modes of transmission and human behaviour – and what governments are and should be doing about it. Newspapers and television channels become much more interested in interviewing academics and highlighting their research. When sharing knowledge, however, consider how academic studies and findings might be received by a non-academic audience. There is much value in producing plain language summaries for journalists and the public [1] to prevent information from being misinterpreted. It is also important to anticipate what questions people are asking each other about the pandemic [36], as these may be different to the ones scientists are most interested in answering.

In April 2020, European researchers from The Netherlands and Belgium caused much concern by sharing findings that suggested aerosolised particles from joggers and cyclists had the potential to spread further than the 1.5m recommended for social distance [51]. The paper (which had not been peer reviewed, but few non-academics fully understand this process) did not suggest that this increased infection risk, simply that some viral particles being carried that far is not an impossibility and therefore warrants mindfulness of distance. The paper was interpreted by media outlets desperate for headline-grabbing stories and by members of the public with existing health anxiety, as meaning that people passed on the street by an unmasked cyclist or jogger were likely to become infected. Professor Bert Blocken, the lead author, admitted he had not expected the attention the paper attracted, nor anticipated that a study on particle aerodynamics would be interpreted as being about infection risk [52]. Similarly, articles showing that the SARS-Cov2 virus has mutated have been sensationalised in the media, with inference that these will be more deadly: to the average member of the public, the term 'mutant' brings forth images of science-fiction monsters, requiring the science journal *Nature* to print an article debunking this view [53]. Explaining, alongside research presentations at the time they are originally discussed, that such evolution is routine and not necessarily cause for alarm can help to dispel misinformation before it takes hold. Before speaking to the press, think like a journalist: how could your work be sensationalised and presented? What conclusions might a member of the public jump to? Make sure, in presenting the research, that such misinterpretations are headed off, particularly those that could lead to actual harm.

The same is true of forward projections – a term deliberately preferred in academic and policy research to 'predictions'. Projections tend to present a range of possible outcomes, the high-end of which is generally the situation that could arise if no mitigation efforts are taken, and which is therefore unlikely to come to pass if the situation is addressed. The lower bound of such predictions may represent the more likely outcome, with the reports in which the data are presented providing the roadmap of how to achieve this. Once again, however, the high estimates from such data are more likely to be sensationalised in the media as what will happen, with the mitigating strategies that should prevent them overlooked. In her report to the UK Government on the country's response to the 2009 H1N1 "Swine Flu" pandemic, Dame Diedre Hine warned that projects do not predict the future, and that this should be borne in mind whenever such predictions are discussed [54].

## 4. Engage with the media to help explain your own and others' findings to the non-academy

Spelling out the ways in which research and scientific data can be misinterpreted by the media and the public may seem counter-intuitive in a paper that is calling for more engagement between scientists and both traditional and social media, but in essence it is a modern-day version of the famous military mantra "know your enemy" [55] – in this case, not the SARS-Cov2 virus, but the channels and mediums through which misinformation about it might spread, perhaps more quickly than the virus itself [56], and the science (il)literacy of the receiving reader. The WHO infodemic management programme starts from the position that interventions and messages must be based on science and evidence, and must reach citizens and enable them to make informed decisions on how to protect themselves and their communities in a health emergency [1]. Scientists and public health professionals therefore have a duty, in times of uncertainty and fast-moving information construction, to help those less able to navigate the information flowing over them, to make complex science clearer, and to take the time and effort to avoid and correct misunderstandings [57].

It would be unfair, however, to suggest that this is not without challenges: taking a stance for science can be not only time consuming but also frustrating. In 2016, Dr Phil Williamson, a marine scientist at the University of East Anglia UK, objected to articles published in the politically-motivated *Breibart News* and *The Spectator* that refuted marine scientists' work on the links between climate change and ocean acidification, He called fellow scientists to not only communicate more widely but also to correct misinformation and errors where necessary [57]. Frustratingly, he lost an official complaint to the press commission and was openly attacked by the *Breitbart* journalist who wrote the original story [58]. Nonetheless, Williamson stands by the importance of speaking up against such anti-science. Increasingly, academic journals are encouraging health professionals and researchers to do the same, particularly against politicians who deny and denigrate science [59, 60]. News agencies such as the UK's Science Media Centre [61] are connecting more responsible media outlets with scientists and researchers who are happy to engage with them, building bridges and encouraging more accurate reporting, from which everyone benefits. But they are dependent on the cooperation of academics and scientists who are prepared to give their time and build relationships with media agencies, journalists and media platforms.

**5. Engage with social media to fill the information vacuum and offer better information**

Academics and scientists who are reluctant to have their words mediated by a journalist can also communicate directly with the public, through social media channels. During COVID19, there have been many successful examples of this, including epidemiologist Dr Trevor Bedford who engages on Twitter [62], Dr John Campbell on YouTube [63], and the moderator team of the r/coronavirus discussion forum on www.reddit.com [64], most of whom are research scientists, medical professionals and academic in fields including virology, epidemiology, paediatrics, public health and biosafety.

Recognising the opportunities social media offers for challenging misconceptions is particularly valuable: misinformation can be challenged directly, providing platform users with better information to challenge things they see elsewhere, while the voting systems on websites such as reddit act as a form of quality control. Previous research has shown such voting mechanisms do tend to improve the quality of information and ensure that the better information rises to the top on platform where information is peer-curated [25].

When engaging, it is important to understand the source of the misinformation and how it has arisen: a concerned citizen who may think they are genuinely uncovering facts others are trying to hide. They may be open to explanations from scientists who understand how novel viruses emerge, biosecurity experts who understand how difficult it would be to manufacture a biological weapon and political scientists who see no immediate motive for doing so [65, 66] and change their stance once this has been explained. It is important for such experts to see the value in debunking rather than simply ignoring or censoring such information. A good example of this is the response to a preprint posted on the academic archiving site *BioRxiv* on 1 Feb 2020 [67], which seemed to imply that the SARS-Cov2 had characteristics suggestive of artificial origin. Speculation on this on the popular online platform reddit [68] was quashed not only by the reddit community itself, but also by the forum's moderator team, a number of whom were research scientists and were able to post a formal response [69]. In an interesting twist, the moderator team recruited two additional moderators from other scientists who engaged with the discussion to debunk the original paper.

Young people in particular may benefit from seeing such discussions played out: they tend to be exposed to a wider range of viewpoints online but do undertake critical fact-checking rather than accepting what they see at face value. Heightened literacy and better understanding of the issues helps such readers to think critically and to make informed decisions on what they read. Knowledge claims are validated between different audiences. Conflicts between populist and expert opinion are unavoidable; the issue is how competing claims are mediated [15]. This can have particular value where misinformation is coming from a source that should be trusted – such as former U.S. President Donald Trump, who was identified as the key source of misinformation online during the COVID19 pandemic in a study by Cornell University [70].

**Conclusions:**

We are living in an interconnected world in which the infodemic recognises no boundaries [1]. Many scientists and academics have already risen to the challenge, making suggestions of how to fight it [71], producing frameworks for action [1], and undergoing WHO training in infodemic management [72]. Leading academics [60] and entire academic journals [73] have taken a political stance against the misinformation spread by Donald Trump, stressing the greater value of Joe Biden's pro-science approach to the pandemic. Whether this had an impact on the U.S. election results or not, it certainly signalled to the U.S. public who the scientists and academics thought was best placed to steer their country through the most serious pandemic of their lifetime. All of us have a similar role to play. All of us are educators. We work with evidence and facts, shape them into knowledge and send that knowledge out into the world where it changes lives, shapes the future and empowers others. As the way knowledge is collected, discussed, shaped, stored, disseminated and accessed evolves, academics and scientists must evolve with it. We must ride the infodemic wave so that, just as with the COVID19 pandemic itself, we stay ahead of, and flatten, its curve.

**Citations**